\documentclass[graybox]{svmult}

\usepackage{type1cm}        
\usepackage{makeidx}         
\usepackage{graphicx}        
\usepackage{epsfig}
\usepackage{multicol}        
\usepackage[bottom]{footmisc}

\usepackage{newtxtext}       %
\usepackage{newtxmath}       

\usepackage{natbib}
\newcommand{\be}{\begin{equation}}
\newcommand{\ee}{\end{equation}}
\newcommand{\bea}{\begin{eqnarray}}
\newcommand{\eea}{\end{eqnarray}}
\newcommand{\bdm}{\begin{displaymath}}
\newcommand{\edm}{\end{displaymath}}

\makeindex             

\begin{document}

\title*{Preferred basis, decoherence and a quantum state of the Universe}

\author{Andrei O. Barvinsky and Alexander Yu. Kamenshchik}

\institute{Andrei O. Barvinsky \at Theory Department, Lebedev Physics Institute, Leninsky Prospect 53, Moscow 119991, Russia\\
  \email{barvin@td.lpi.ru}\\
  Alexander Yu. Kamenshchik \at Dipartimento di Fisica e Astronomia,
Universit\`a di Bologna and INFN, via Irnerio 46, 40126 Bologna, Italy\\
Landau Institute for Theoretical Physics, Russian Academy of Sciences, Kosygin street 2, 119334 Moscow, Russia\\
\email{kamenshchik@bo.infn.it}}

\maketitle

\abstract{
We review a number of issues in foundations of quantum theory and quantum cosmology including, in particular, the problem of the preferred basis in the many-worlds interpretation of quantum mechanics, the relation between this interpretation and the decoherence phenomenon, application of decoherence approach to quantum cosmology, the relation between the many-worlds interpretation and Anthropic Principle along with the notion of quantum-classical duality. We also discuss the concept of fundamentally mixed quantum state of the Universe represented by a special microcanonical density matrix and its dynamical realization in the form of the semiclassically treated path integral over spacetime geometries and quantum matter fields.
These issues can be considered as a part of the scientific legacy of H. D. Zeh generously left to us in his two seminal papers published at the beginning of seventies in Foundations of Physics. }

\section{Introduction}

As is well-known  quantum mechanics and more generally quantum theory (including quantum statistical mechanics and
quantum field theory) have had great achievements in the description of microworld. Its arrival not only has brought
a lot of new technologies but also has changed our vision of the world. Since the days of its creation the problems of interpretation of quantum mechanics have been attracting attention of a
 person working in this field as well as  of a broader public, including philosophers, psychologists, biologists and even of the people of arts and literature (see e.g. \cite{Jammer}). The main feature of quantum mechanics, which distinguishes it from classical Newton mechanics is the fact, that even if one has a complete knowledge of a state of a system under consideration and would like to make a certain experiment, more than one
alternative result of such an experiment is possible. The knowledge of the state of the system can permit us only to calculate the probabilities of different outcomes of the experiment, as was first understood by Max Born( \cite{Born}).
However, a natural question arises how can we see only one outcome of an experiment and what happens with all other alternatives? One can present this question in a slightly different form: how can we reconcile ourselves with the fact that the physics of the microworld is described by quantum mechanics while in the macroworld, which we perceive in our everyday experience we encounter  the laws of classical physics?

At the dawn of quantum mechanics the very existence of different alternatives and non-classicality of microworld was a source of some kind of dismay.   The first attempt to cope with this situation was undertaken in the framework of the so called Copenhagen interpretation of quantum mechanics, which represents a collection of views and ideas of some of the founders of quantum mechanics. In particular, Niels Bohr have suggested the notion of complementarity between different notions and approaches, which seemed to resolve apparent contradictions of the quantum theory. Besides, he insisted on the existence of the so called classical realm, where all the results of experiments and observations were registered. Thus, the classical physics was considered not only as a limiting case of the quantum physics, but also as a pre-requisite of its very existence (\cite{Bohr}).
Finally, it was von Neumann, who formulated the mathematically rigorous idea of the reduction of the wave packet and in such a way had given a constructive picture of events, occurring in the process of quantum measurement (\cite{Neumann}). According to von Neumann, in quantum mechanics there coexisted two processes. One of them is a unitary deterministic evolution of the wave function, describing the quantum system, according to the Schr\"odinger equation. The second process takes place during quantum measurement and is called  the reduction of the wave function, when one of the possible outcomes is realized, while others disappear into thin air. In the process of quantum measurement
three players participate: an object, a measuring device and an observer, and the presence of the latter two corresponds in a way to the Bohr's  classical realm. Thus, in this picture, everything which one calculates and predicts in quantum theory finds its explanation.
Generally, one can say that according  to the Copenhagen interpretation the classical and quantum mechanics were present in our physical world on equal footing. The Copenhagen interpretation has played a very important role in the development of quantum mechanics and its applications.  It has made the physicists' imagination more free and have made them more accustomed to the idea that the deterministic ideal of classical mechanics is not an absolute goal of the physical theory.
However, some other founders of quantum theory such as Planck, Einstein, Schr\"odinger and de Broglie were not happy with the Copenhagen interpretation and thought that some rebirth of the classical ideal was necessary. The most consistent attempt of such a rebirth was undertaken by D. Bohm and is known as de Broglie-Bohm interpretation (\cite{Bohm}).

On the other hand, the presence of two dynamical processes in quantum theory looked logically unsatisfactory
and in 1957  Hugh Everett has published a short version of his PhD thesis  (\cite{Everett}) under the title
``Relative-state formulation of quantum mechanics''. It contained a simple idea. We do not need the postulate of the reduction of the wave packet and hence, only one fundamental process exists in quantum theory -- unitary evolution
governed by the Schr\"odinger equation. All the outcomes of the experiment co-exist and the objective result of the measurement under consideration is the establishment of correlations between the measured and measuring subsystems, which are treated on equal footing. Thus, there is no need in a special classical realm too.

The Everett interpretation of quantum mechanics seemed to be quite logical and economical, however, this economy was achieved by means of the acception  of parallel existence of different outcomes of a quantum measurement and this was a critical point. Indeed, behind the mask of the relative states of two or more subsystems loomed a disturbing image of the co-existence of parallel worlds and of the splitting reality. Probably this fact explains a rather troubled history of the Everett interpretation recognition by the
scientific community (\cite{Byrne}).  In 1970 B. S. DeWitt  published the paper in Physics Today (\cite{DeWitt1}) and then edited a book (\cite{DeWitt}), which have given a new birth to the Everett interpretation of quantum mechanics under the name ``many-worlds'' interpretation.

For many years the many-worlds interpretation  was treated as something rather exotic.  However, now the situation is changing due to two main developments in
quantum physics: progress in the study of  quantum cosmology  (see e.g. \cite{Kiefer})
and the birth of quantum informatics (see e.g. \cite{Nielsen}).
In quantum cosmology one treats the universe as a unique quantum object. Thus, there is no
place for a classical external observer and other agents, who could be responsible for the presumed reduction of the wave function. Quantum algorithms, in their turn, use essentially the parallel quantum processes, which also marginalises  the idea of a quantum realm.

Generally, one can say that the idea that quantum theory is more fundamental than classical one becomes more and more popular. However, an important question remains: why in many situations we see the classical behaviour of macroscopic objects?
What are the relations between  classical and quantum theories? Heinz - Dieter Zeh has made an important contributions in this field. In the beginning of the 70-th he has published two seminal papers in Foundations of Physics. In one of these papers
(\cite{measurement}) he has open a new direction - the study of decoherence. In another paper (\cite{observation}) he considered the problem of the choice of the preferred basis in the many-worlds interpretation of quantum mechanics. From our point of view these two approaches complement each other. In this paper we present a review of several works whose content and results are closely related to the works of H. D. Zeh. 

The structure of the paper is the following: in the second section we discuss the problem of the preferred basis in quantum mechanics; in the third section we describe the dynamics of the preferred basis, while the fourth section is devoted to the interrelations between the phenomenon of decoherence and the many-worlds approaches. In the fifth section we discuss applications of the decoherence approach to cosmology in connection with the problem of ultraviolet divergences in quantum gravity; the sixth section is devoted to the notion of quantum-classical duality; in the seventh section we consider the relations between the many-worlds interpretation and Anthropic Principle; the eighth section is devoted to the cosmological model in which the notion of the density matrix of the universe arises; the ninth section contains brief concluding remarks.

\section{The problem of the preferred basis in the many-worlds interpretation of quantum mechanics}

As we have already mentioned, in the framework of many-worlds interpretation
Schr\"odinger evolution is the only process. Thus,
 the principle of superposition is applicable to all the states including macroscopic ones and all the
outcomes of any measurement-like processes are realized simultaneously but in different ``parallel universes''.
The very essence of the many-worlds interpretation can be expressed by one simple formula.
Let us consider the wave function of a system, containing two subsystems (say, an object and a device),
whose wave functions are respectively $|\Phi\rangle$ and $\Psi\rangle$ and let the process of interaction between these two subsystems be described by a unitary operator $\hat{U}$. The result of the action of this operator can be represented as
\begin{equation}
\hat{U}\,|\Phi\rangle_0 \Psi\rangle_{i} = |\Phi\rangle_i \Psi\rangle_{i}.
\label{unitary}
\end{equation}
Here the state $|\Psi\rangle_{i}$ is a quantum state of the object corresponding to a definite outcome of the
experiment, while $|\Phi\rangle_0$ is an initial state of the measuring device. Now, let the initial state
of the object be described by a superposition of quantum states:
\begin{equation}
|\Psi\rangle = \sum_{i} c_i |\Psi\rangle_i.
\label{superposition}
\end{equation}
Than the superposition principle immediately leads to
\begin{equation}
\hat{U}|\Phi\rangle_0 \Psi\rangle = \hat{U}|\Phi\rangle_0\sum_{i} c_i |\Psi\rangle_i =
\sum_{i}c_i|\Phi\rangle_i \Psi\rangle_i.
\label{superposition1}
\end{equation}
Here $|\Phi\rangle_i$ describes the state of the measuring device, which has found the quantum object in the
state $|\Psi\rangle_i$. The superposition  (\ref{superposition1}) contains more than one term, while one
sees only one outcome of measurement. The reduction of the wave packet postulate solves this puzzle by
introducing another process eliminating in a non-deterministic way all the terms in the right-hand side
of Eq. (\ref{superposition1}) but one. All the terms  of the superposition
are realized but in different universes.

Here, inevitable question arises:
decomposing the wave function of the universe one should choose a certain basis. The result of
the decomposition essentially depends on it. Thus, the so called  problem of the choice of the preferred basis arises.
This problem was considered by different authors (see e.g. (\cite{Deutsch}), (\cite{Markov}), (\cite{Dieks}), (\cite{BenDov}), (\cite{Albrecht})).
From our point of view the most convincing approach to the solution of this problem was proposed by Zeh in \cite{observation}
and further developed in our works (\cite{Barv1}, \cite{Barv3}, \cite{Barv4}, \cite{Barv5}).
The essence of the problem can be formulated considering
the same example of a quantum system consisting of two subsystems. Let us emphasize that now we would like
to undertake a consideration of a general case without particular reference to measuring devices
and quantum objects (for a moment we consider this division of a system into subsystems as granted).
The only essential characteristics of the branching process is the defactorization of the wave function.
This means that if at the initial moment the wave function of the system under consideration was represented
by a direct product of the wave functions of the subsystems
\begin{equation}
|\Psi\rangle = |\phi\rangle |\chi\rangle
\label{direct}
\end{equation}
then after an interaction between the subsystems it becomes
\begin{equation}
\sum_{i} c_i |\phi\rangle_i |\chi\rangle_i,
\label{defact}
\end{equation}
where more than one coefficient $c_i$ is different from zero.
Apparently the decomposition (\ref{defact}) can be done in various manners. As soon as each term
is associated with a separate universe, the unique prescription for the construction of such a
superposition should be fixed.  We believe that the correct choice of the preferred basis is the  so
called Schmidt (\cite{Schmidt}) or bi-orthogonal basis. This basis is formed by eigenvectors of both the density matrices of the subsystems of the quantum system under consideration.
These density matrices are defined as
\begin{equation}
\hat{\rho}_{I} = Tr_{II}|\Psi\rangle \langle \Psi|,
\label{density}
\end{equation}
\begin{equation}
\hat{\rho}_{II} = Tr_{I}|\Psi\rangle \langle \Psi|.
\label{density1}
\end{equation}
Remarkably, the eigenvalues of the density matrices coincide and hence the number of non-zero eigenvalues is the same, in spite of the fact that the corresponding Hilbert spaces can be very different.
\begin{equation}
\hat{\rho}_{I} |\phi_n\rangle = \lambda_n |\phi_n\rangle,
\label{density2}
\end{equation}
\begin{equation}
\hat{\rho}_{II} |\chi_n\rangle = \lambda_n |\chi_n\rangle,
\label{density3}
\end{equation}
Consequently, the wave function is decomposed as
\begin{equation}
|\Psi\rangle = \sum_{n} \sqrt{\lambda_{n}}|\phi_n\rangle|\chi_n\rangle.
\label{density4}
\end{equation}

The bi-orthogonal basis
was first used at the dawn of quantum mechanics by E. Schr\"odinger (\cite{Schrodinger1,Schrodinger2})  for the study of correlations between
quantum systems. Recently, this basis has been actively used for measuring the degree of entanglement, in particular, in relation
to quantum computing (\cite{Ekert}).

We believe that the bi-orthogonal basis being defined by the fixing of the decomposition of the system
into subsystems has a fundamental character and determines the worlds which result from the defactorization process. However, the subdivision of the system into subsystems which implies the branching of the worlds
should satisfy some reasonable criteria which we are not ready to formalize at the moment (see, however our paper (\cite{Barv5}) for the analysis of some relatively simple cases). One can say, that the decomposition into the subsystems should be such that the corresponding preferred basis would be rather stable. For example, when one treats a quantum mechanical experiment of the Stern-Gerlach type, it is natural to consider the measuring device and the atom as subsystems.   In the case when we consider a system with some kind of internal symmetry, like in quantum chromodynamics, the division of the system into subsystems which belong to singlet representations of the internal symmetry group looks also reasonable from the point of view of stability of the bi-orthogonal preferred basis of the many-worlds interpretation.

\section{Dynamics of the preferred basis}

Let us consider a system, consisting of two subsystems. In the preferred basis its wave function has the form
\begin{equation}
\Psi\rangle = \sum_nc_n|n\rangle_{I}|n\rangle_{II}.
\label{dyn}
\end{equation}
The whole wave function $|\Psi(t)\rangle$ satisfies the Schr\"odinger equation
\begin{equation}
i\frac{\partial|\Psi(t)\rangle}{\partial t} = H|\Psi(t)\rangle,
\label{dyn1}
\end{equation}
with
\begin{equation}
H = H_{I}+H_{II}+V,
\label{dyn2}
\end{equation}
where $V$ is the interaction Hamiltonian between the subsystems.
One can show (\cite{Barv3,Barv4}) that the evolution of the preferred basis vectors is unitary and is governed by some effective Hamiltonians:
\begin{equation}
i\frac{\partial}{\partial t}|m\rangle_{I} = {\cal H}_{I}|m\rangle_{I},
\label{dyn3}
\end{equation}
\begin{equation}
{\cal H}_{I} = i\sum_m\left(\frac{\partial |m\rangle_{I}}{\partial t}\   _I\langle m|\right).
\label{dyn4}
\end{equation}
(The equations for the second subsystem are quite analogous).
Non-diagonal elements of ${\cal H}_I$ are given by
\begin{equation}
({\cal H_I})_{mn}=(H_I)_{mn}-\frac{_I\langle m|{\rm Tr}_{II}[V,|\Psi\rangle\langle\Psi|]|n\rangle_I}{p_m-p_n},
\label{dyn5}
\end{equation}
and, similarly, for ${\cal H}_{II}$. The diagonal elements can be fixed as
\begin{equation}
({\cal H}_{I})_{mm}=(H_I)_{mm};\ ({\cal H}_{II})_{mm}=(H_{II})_{mm},
\label{dyn6}
\end{equation}
whence it follows that
\begin{equation}
i\frac{\partial c_n}{\partial t} =\, _I\!\langle n|_{II}\langle n|V|\Psi\rangle.
\label{dyn7}
\end{equation}
Equations \eqref{dyn5} and \eqref{dyn6} completely determine the effective Hamiltonians which turn out to be a complicated functionals not only of the hamiltonian $H$, but also of the quantum state $|\Psi\rangle$ of the system.
A similar consideration can be found in the paper by K\"ubler and Zeh (\cite{dynamics}).

The most unexpected conclusion from the unitary dynamics of the proposed basis is the following. Since the observer (identified, for example, with the subsystem $I$) observes and measures only one relative state of the second subsystem $II$ in his many-worlds branch, he finds that this state undergoes a unitary evolutions of the above type. This is in spite of the impure nature of this open subsystem $II$ described by a non-factorizable density matrix.

The second conclusion is that the observer studying the dynamics of his relative state measures the effective Hamiltonian
${\cal H}_{II}$ and not the fundamental Hamiltonian $H$ of the total system. This apparently means that research into nature at the most fundamental levels requires additional efforts in reconstructing the fundamental dynamical laws on the grounds of the observable reality. Other aspects of the relations between observable and fundamental Hamiltonians which we call ``quantum-classical duality'' were considered in (\cite{Kamenshchik}) and will be briefly described in section 6.

\section{Preferred basis and decoherence}

In his seminal paper (\cite{measurement}) Heinz-Dieter Zeh has open a new direction in quantum theory -- decoherence. Let us briefly remind what is it. Let us consider for example the Stern-Gerlach experiment, where the initial state of the atom and of the device is described by the vector
\begin{equation}
|\Psi\rangle = |\Phi\rangle(c_1|\psi_{\uparrow}\rangle+c_2|\psi_{\downarrow}\rangle).
\label{dec}
\end{equation}
As a result of the measurement, we have
\begin{equation}
|\Psi\rangle_{\rm final} = c_1|\Phi\rangle_{\uparrow}|\psi_{\uparrow}\rangle+c_2|\Phi\rangle_{\downarrow}|\psi_{\downarrow}\rangle.
\label{dec1}
\end{equation}
Let us remember that in every measurement process the third system participates - it is the environment. Hence, the initial state of the general system is
\begin{equation}
|\Psi\rangle = |\chi\rangle|\Phi\rangle(c_1|\psi_{\uparrow}\rangle+c_2|\psi_{\downarrow}\rangle),
\label{dec2}
\end{equation}
while its final state is
\begin{equation}
|\Psi\rangle_{\rm final} = c_1|\chi\rangle|_{\uparrow}|\Phi\rangle_{\uparrow}|\psi_{\uparrow}\rangle+c_2|\chi\rangle_{\downarrow}|\Phi\rangle_{\downarrow}|\psi_{\downarrow}\rangle.
\label{dec3}
\end{equation}
Now, if we calculate the reduced density matrix tracing out the environmental degrees of freedom, we obtain
\begin{equation}
\rho_{\rm reduced}={\rm Tr}_{\{\chi\}}|\Psi\rangle_{\rm final\  final}\langle \Psi|=|c_1|^2|\Phi_{\uparrow}\rangle\langle\Phi_{\uparrow}|
\psi_{\uparrow}\rangle\langle\psi_{\uparrow}|+ |c_2|^2|\Phi_{\downarrow}\rangle\langle\Phi_{\downarrow}|
\psi_{\downarrow}\rangle\langle\psi_{\downarrow}|.
\label{dec4}
\end{equation}
The expression \eqref{dec4} represents a classical statistical mixture, which substitutes the quantum state due to tracing out the environmental degrees of freedom.  This is the essence of the decoherence approach. It explains how the classical world arises from the quantum one and has proved its efficiency for description of a very wide range of phenomena in theoretical and experimental physics (see \cite{deco}).  Thus, the idea of application of the decoherence approach to quantum cosmology was quite natural, and such applications were studied using the minisuperspace models (\cite{Kiefer1, Kiefer2}). In the next section we shall discuss the problem arising in cosmological models with infinite number of degrees of freedom.

Before concluding this section we would like to make several remarks about the comparison between
the many-worlds approach to the interpretation of quantum mechanics and the decoherence approach.  As we have already told, what lies in the foundation of the decoherence  approach is understanding the fact that in any process of quantum measurement there are not two, but three participants: namely, not only the quantum object and measuring device, but also the rest
of the universe - the so called environment. After the measurement, we can construct
the reduced density matrix, describing the object and device, tracing out unobservable degrees of freedom of the environment. It appears that in many cases this reduced
density matrix becomes quickly practically diagonal in a certain ``good'' basis, whose states are sometimes called ``pointer states'' (\cite{Zurek1,Zurek2}) and behaves more or less classically. In such a way, the quantum state of the object and of the measuring device
becomes a classical statistical mixture.  However, from our point of view
the decoherence approach to the problem of quantum measurement and to the problem
of classical-quantum relations is less fundamental than the many-worlds approach.

First, there is an essential difference between statistical principles in classical and quantum physics. In classical physics the probability is  ``the measure of our ignorance'' of the initial conditions or of the details of interaction while in quantum physics we cannot get rid of the probability even in principle, where there is no analogue to the ``Laplace demon'', who can calculate everything. Thus, the transition to a classical statistical mixture does not resolve the problem of choice between different alternatives.

Second, the decoherence properties of reduced density matrix depend crucially on the choice
of the basis. Thus, the classicality is introduced into the theory already at the level of the choice of the basis. In the bi-orthogonal preferred basis approach, described above, the
basis is defined by the chosen decomposition of the system under consideration into subsystems. After that, one can study the dynamics of different elements of the basis
and to see if they behave classically (\cite{Barv4}). It appears, that
sometimes classicality exists as a stable phenomenon, sometimes -- as a temporary phenomenon and sometimes it does not exist at all. Thus, the many-worlds interpretation,
insisting on  the primary role of the quantum theory with respect to the classical one,
describes a wider class of phenomena. Nevertheless, for a large class of situations,
the predictions of both approaches are close. It happens when the bi-orthogonal basis
is close to the pointer basis.

\section{Decoherence and ultraviolet divergences in quantum cosmology}

The first question which arises when one tries
to explain the classicalization of a quantum Universe
using decoherence approach, is connected with
the definition of an environment. Indeed, in contrast
to the usual description of quantum-mechanical
experiment in a laboratory, there is no external
environment, because the object of quantum cosmology
is the whole Universe. Thus, we should
treat some part of degrees of freedom as essential
and observables, while the others could be treated as
an environment with subsequent tracing them out
in transition to a reduced density matrix.
It is natural to believe that inhomogeneous degrees
of freedom play the role of environment
while macroscopic variables such as a cosmological
radius or initial value of the inflaton scalar field
are treated as observables. It is easy to guess that
there are infinite number of environmental degrees
of freedom and hence, calculating a reduced
density matrix encounters the problem of
ultraviolet divergences, which was analyzed in papers  (\cite{Barv11, Barv12, Barv13}).

To tackle this problem we should consider the
wave function of the universe in the one-loop approximation,
describing simultaneously with homogeneous
also inhomogeneous degrees of freedom (\cite{Barv8, Barv9, Barv10}).
It is convenient to write both the no-boundary (\cite{Hawk}) and tunneling
(\cite{Vilenkin})
cosmological wave functions in the form:
\begin{equation}
\Psi(t|\varphi,f)=\frac{1}{\sqrt{v^*_{\varphi}(t)}}\exp\Big(\mp I(\varphi)/2+iS(t,\varphi)\Big)\times\prod_n\psi_n(t,\varphi|f_n),
\label{dec-cosm}
\end{equation}
\begin{equation}
\psi_n(t,\varphi|f_n)=\frac{1}{\sqrt{v_n^*(t)}}\exp\left(-\frac12\Omega_n(t)f_n^2\right),
\label{dec-cosm1}
\end{equation}
\begin{equation}
\Omega_n(t)=-a^k(t)\frac{\dot{v}_n^*(t)}{v_n^*(t)}.
\label{dec-cosm2}
\end{equation}
Here, the sign minus or plus in front of Euclidean
action $I(\varphi)$  in the exponential of \eqref{dec-cosm} corresponds to
the no-boundary  and to the tunneling  wave
functions of the Universe, respectively, $f_n$ describe
amplitudes of inhomogeneous modes, while $v_n$
 are corresponding to these modes solutions
of linearized second-order differential equations.
The exponential $k$ in the expression for the frequency
function $\Omega_n$, depends on the spin $s$ of the
field under consideration and on its parametrization.
For the ``standard'' parametrization $k =
3 - 2s$. Inclusion of inhomogeneous modes into
the wave function of the Universe were first considered
in (\cite{Halliwell, Laflamme}).
One can show (\cite{Barv8}) that the diagonal of the reduced density matrix
corresponding to the wave function \eqref{dec-cosm}
\begin{equation}
\rho(t|\varphi)\equiv\rho(t|\varphi,\varphi) = \int\prod_n df_n|\,\Psi(t|\varphi,f)\,|^2
\label{dec-cosm3}
\end{equation}
is
\begin{equation}
\rho(t|\varphi) = \frac{\sqrt{\Delta_{\varphi}}}{|v_{\varphi}(t)|}\exp\Big(\!\mp I(\varphi)-\varGamma_{\rm 1-loop}(\varphi)\Big),
\label{dec-cosm4}
\end{equation}
where
\begin{equation}
\Delta_{\varphi}\equiv ia^k(v_{\varphi}^*\dot{v}_{\varphi}-\dot{v}_{\varphi}^*v_{\varphi}).
\label{dec-cosm5}
\end{equation}
is the Wronskian of the $\varphi$-mode functions and $\varGamma_{\rm 1-loop}$  is the one-loop effective action
calculated on the DeSitter instanton of the
radius $1/H(\varphi)$, where $H(\varphi)$ is the effective Hubble constant.
When $H(\varphi) \rightarrow \infty$,
\begin{equation}
\varGamma_{\rm 1-loop} = Z\ln\frac{H(\varphi)}{\mu},
\label{dec-cosm6}
\end{equation}
where $Z$ is the anomalous scaling of the theory, $\mu$
is a renormalization scale. It is easy to see
that the condition of the normalizability of the
wave function of the Universe is
\begin{equation}
Z > 1,
\label{dec-cosm7}
\end{equation}
and this condition provides us with the selection
criterium for particle physics models (\cite{Barv8}).

Information about decoherence behaviour
of the system is contained in the off-diagonal elements
of the density matrix. In our case they read
\begin{equation}
\rho(t|\varphi,\varphi')=
\left(\frac{\Delta_{\varphi}\Delta_{\varphi'}}{v_{\varphi}v_{\varphi'}^*}\right)^{\frac14}
\exp\left(-\frac12\varGamma-\frac12\varGamma'+i(S-S')\right)D(t|\varphi.\varphi').
\label{dec-cosm8}
\end{equation}
Here $D(t|\varphi,\varphi')$ is the so called decoherence factor
defined by the formula
\begin{equation}
D(t|\varphi,\varphi')=\prod_n\left(\frac{4Re\Omega_n Re\Omega_n'^*}{(\Omega_n+\Omega_n'^*)^2}\right)^{\frac14}\left(\frac{v_nv_n'^*}{v_n^*v_n'}\right)^{\frac14}.
\label{dec-cosm9}
\end{equation}
How to cope with ultraviolet divergences appearing
in the sum of this type? This qiestion
was already discussed in (\cite{ultra, ultra1, ultra2}). In (\cite{Barv11, Barv12, Barv13}) we  used the dimensional regularization (\cite{dimen}).
As usual, the main effect of a dimensional
regularization consists in changing the number of degrees of freedom involved in summation. For example, for a scalar field, the
degeneracy number of harmonics in spacetime
of dimensionality $d$ changes from the well-known value (see e.g. \cite{Khalat})
\begin{equation}
{\rm dim}(n, 4) = n^2,
\label{dec-cosm10}
\end{equation}
to
\begin{equation}
{\rm dim}(n, d) = \frac{(2n+d-4)\varGamma(n+d-3)}{\varGamma(n)\varGamma(d-1)}.
\label{dec-cosm11}
\end{equation}
Making analytical continuation and discarding
the poles $1/(d-4)$ one has finite values
for $D(t|\varphi,\varphi')$. However, for scalar, photon and
graviton fields one gets because of oversubtraction of UV-infinities a pathological behaviour:
\begin{equation}
|D(t|\varphi,\varphi')| \rightarrow \infty,\ {\rm at}\ |\varphi-\varphi'| \rightarrow \infty.
\label{dec-cosm12}
\end{equation}
For example, for a massive scalar field
\begin{equation}
\ln|D(t|\varphi,\varphi')| \approx \frac{7}{64}m^3\bar{a}(a-a')^2,
\label{dec-cosm13}
\end{equation}
where
\begin{eqnarray}
&&a=\frac{1}{H(\varphi)}\cosh H(\varphi)t,
\nonumber \\
&&a'=\frac{1}{H(\varphi')}\cosh H(\varphi')t,
\nonumber \\
&&\bar{a}=\frac{a+a'}{2}.
\label{dec-cosm14}
\end{eqnarray}
Such a form of a decoherence factor not only does
not correspond to decoherence, but also renders
density matrix ill-defined, breaking the condition
${\rm Tr}(\rho^2) \leq 1$.
However, there is remedy: using
the reparametrization of a bosonic scalar field
\begin{equation}
f \rightarrow \tilde{f} = a^{\mu}f,\ v_{n} \rightarrow \tilde{v}_n=a^{\mu}v_n,
\label{dec-cosm15}
\end{equation}
one can get the new form of the frequency function
\begin{equation}
\Omega_n(t)=-ia^{3-2\mu}(t)\frac{\dot{\tilde{v}}_n^*(t)}{\tilde{v}_n^*(t)}.
\label{dec-cosm16}
\end{equation}
In such a way one can suppress ultraviolet divergences.
For the so called conformal parametrization,
$\mu = 1$, for the massive scalar field one has
\begin{equation}
\ln|\tilde{D}(t|\varphi,\varphi')|=-\frac{m^3\pi\bar{a}(a-a')^2}{64},
\label{dec-cosm17}
\end{equation}

For the case of fermions this trick does not work (\cite{Barv12}). Let us discuss it in more detail. The wave function of the Universe filled
by fermions has the form (\cite{D'Eath, Kiefer-ferm, Barv12})
\begin{equation}
\Psi(t,\varphi|x,y) = \Psi_0(t,\varphi)\prod_n\psi_n(t|x_n,y_n),
\label{dec-cosm18}
\end{equation}
where $x, y$ - Grassmann variables.
Partial wave functions have the form
\begin{equation}
\psi_n(t|x_n,y_n)=v_n-\frac{i\dot{v}_n+\nu v_n}{m}x_ny_n,
\label{dec-cosm19}
\end{equation}
where the functions $v_n$ satisfy the second-order equation
\begin{equation}
\ddot{v}_n+(-i\dot{\nu}+m^2+\nu^2)v_n=0,\quad \nu = \frac{n+\frac12}{a}.
\label{dec-cosm20}
\end{equation}
As was shown in (\cite{Kiefer-ferm})
\begin{equation}
|D(a,\varphi|a',\varphi')|=\exp\left(-\frac{m^2(a-a')^2}{8}\sum_{n=1}\frac{n(n+1)}{\left(n+\frac12\right)^2}\right).
\label{dec-cosm21}
\end{equation}
One can try eliminating ultraviolet divergences
here by dimensional regularization using
the fact that for spinors in spacetime of dimensionahty
$d$:
\begin{equation}
{\rm dim}(n,d)=\frac{\varGamma(n+2^{(d-2)})\varGamma(n+2^{(d-2)/2}-1)}{[\varGamma(2^{(d-2)/2})]^2\varGamma(n+1)\varGamma(n)}.
\label{dec-cosm22}
\end{equation}
However, as the result of the renormalization procedure
we have
\begin{equation}
|D(a,\varphi|a',\varphi')|=\exp\left(-\frac{m^2(a-a')^2}{8}I\right),\ I < 0,
\label{dec-cosm23}
\end{equation}
and we enconter the same problem
as in the case of bosons. Moreover, one cannot
use the conformal reparametrization in this case
because standard fermion variables are already
presented in the conformal parametrization.
However, there is another way to circumvent this problem.
One can perform a non-local Bogoliubov transformation
mixing Grassmann variables $x$ and $y$.
This transformation modifies the effective mass of
fermions in equation (\ref{dec-cosm20}) for their basis functions. 
Choosing it in a certain way one can
suppress ultraviolet divergences. The reasonable
idea is to fix this transformation by the requirement that decoherence is absent in static spacetime.
Then:
\begin{equation}
|D(a,\varphi|a',\varphi')|=\exp\left(-\frac{\pi^2m^2(a-a')^2}{192}I\right), \quad I>0,
\label{dec-cosm24}
\end{equation}
and is finite.
The main conclusion to be drawn from the above examples is the
fact that consistency of the reduced density matrix might determine 
the very definition of the environment in quantum cosmology.

\section{Classical - quantum duality}

We have already expressed our opinion that quantum theory is more fundamental than the classical one and that classicality arises under some particular circumstances. In (\cite{Barv4}) we considered toy models
with the Gaussian quantum states and simple interaction Hamiltonians between two subsystems. These Hamiltonians have the structure similar to $V = xp_y+yp_x$ and represent a generalization of the Hamiltonian $xp_y$, introduced by von Neumann
(\cite{Neumann}) to describe the process of quantum measurement. The study of these models confirmed our hypothesis about possible transient character of classicality.

Here we would like to dwell on other aspects of quantum - classical relations following a recent paper (\cite{Kamenshchik}).
Speaking about the problem of time in quantum cosmology and generally, in quantum mechanics, one can remember that some analogue of the classical time can be introduced even in the system with  one degree of freedom (\cite{Sommerfeld, Rowe}). The idea is very simple.  Let us consider a particle with one spatial coordinate and a stable probability distribution for this coordinate. Naturally, the quantum state of such a particle is  an eigenstate of its Hamiltonian. Then, one can suppose that behind this probability distribution there is a classical motion which we can observe stroboscopically. That means that we can detect its position many times and obtain a probability distribution for this position. Classically
 this measured probability is inversely proportional to the velocity of the particle. Indeed, the higher is the velocity of a particle in some region of the space the less is the time that it spends there.
 However in  quantum mechanics this probability is given by the squared modulus of its wave function. Thus,  we can write down the following equality (\cite{Sommerfeld,Rowe})
\begin{equation}
\psi^*(x)\psi(x) = \frac{1}{|v(x)|T},
\label{classic}
\end{equation}
where $T$ is a some kind of normalising time scale, for example, a half period of the motion of the particle.  In this spirit, in  paper (\cite{Rowe}), the probability distributions for the energy eigenstates of the hydrogen atom with a large principal quantum number $n$ were studied. It was shown that the distributions with the orbital quantum number $l$ having the maximal possible value $l=n-1$, being interpreted as in Eq. (\ref{classic}), describe the corresponding classical motion  of the electron on the circular orbit. At the same time, the state with $l=0$ cannot produce immediately a correct classical limit (\cite{Rowe}).  To arrive to such a limit, which represents a classical radial motion of a particle
(i.e.  on a degenerate ellipse) one should apply a coarse-graining procedure based on the Riemann - Lebesgue theorem.
 There is another interesting example: the harmonic oscillator with a large value of the quantum number $n$. In this case, making a coarse-graining of the probability density one can again reproduce a classical motion of the oscillator (\cite{Pauling}).

Usually, when one studies  the question of the classical-quantum correspondence, one looks for the situations where this correspondence is realised. However, it is reasonable to suppose that such situations are not always realised.
Now we would like to attract  attention to another phenomenon: a particular quantum-classical duality between the systems governed by different Hamiltonians (\cite{Kamenshchik}). We can consider  a simple example. Let us suppose that we have a classical motion of the harmonic oscillator, governed by the law
\begin{equation}
x(t) = x_{0}\sin\omega t.
\label{harmon1}
\end{equation}
The velocity is
\begin{equation}
\dot{x}(t)=\omega x_0\cos\omega t.
\label{velocity1}
\end{equation}
Using Eq. (\ref{classic}), we can believe that behind this classical motion there is a stationary wave function
\begin{equation}
\psi(x) = \frac{1}{\sqrt{\pi}(x_0^2-x^2)^{1/4}}e^{if(x)}\theta(x_0^2-x^2),
\label{function1}
\end{equation}
where $\theta$ is the Heaviside theta-function and $f$ is a real function.  Now, applying the energy conservation law and the stationary Schr\"odinger equation we can
find the corresponding potential for the quantum problem:
\begin{equation}V(x) = \frac{m\omega^2x_0^2}{2}+\frac{\hbar^2}{2m}\left(\frac{1}{2(x_0^2-x^2)}+\frac{5x^2}{4(x_0^2-x^2)^2}+if''+if'\frac{x}{x_0^2-x^2}-f'^2\right),
\ {\rm if}\ x^2 < x_0^2,
\label{potential1}
\end{equation}
Here, ``prime'' means the derivative with respect to $x$. To guarantee the reality of the potential and, hence, the hermiticity of the Hamiltonian, we must choose the phase function $f$ such that
\begin{equation}
f' = C\sqrt{x_0^2-x^2},
\label{phase}
\end{equation}
where $C$ is a real constant.
Then the potential (\ref{potential1}) is equal to
\begin{eqnarray}
&&V(x) = \frac{m\omega^2x_0^2}{2}+\frac{\hbar^2}{2m}\left(\frac{1}{2(x_0^2-x^2)}+\frac{5x^2}{4(x_0^2-x^2)^2}+C^2(x^2-x_0^2)\right),\nonumber \\
&&\ {\rm if}\ x^2 < x_0^2.
\label{potential2}
\end{eqnarray}
Then for  $x^2>x_0^2$ we can treat the value of the potential as infinite since there the wave function is zero.

Naturally, the example constructed above is rather artificial.
We have elaborated it to  hint at the possibility of encountering a similar effect in cosmology. One
can imagine  a situation where behind the visible classical evolution of the universe looms a quantum system, whose Hamiltonian  is quite   different
from the classical Hamiltonian  governing this visible classical evolution.

\section{Many-worlds interpretation, probabilities and Anthropic Principle}

As is well known one of the main features of quantum mechanics is the
probabilistic character of its predictions. During many years the Born rule, connecting the probabilities of the outcomes of a quantum measurement with the squared modules of the coefficients  of expansion of the wave function of a system under consideration with respect
to the eigenvectors of the operator, representing the measured quantity, was considered
as a fundamental postulate. However, later the so called Finkelstein - Hartle - Graham theorem
was proven (\cite{Finkelstein,  Hartle,  Graham}). The message of this theorem
consists in the fact that the Born rule can be derived from other postulates of quantum mechanics if one considers a huge number of identical quantum systems and defines the probability as a relative frequency of a chosen outcome of an experiment with respect to
the general number of trials. Finkelstein and Hartle considered this statement without connection with the many-worlds interpretation of quantum mechanics, while Graham
worked in the framework of this interpretation. It seems to us that the Finkelstein - Hartle - Graham theorem looks especially harmonious in the framework of the many-worlds interpretation.

At the same time, the probability treatment of the predictions of quantum mechanics is quite natural when one speaks about
multiple experiments or about multiple identical systems as it happens in the proof of the Finkelstein - Hartle - Graham theorem.
What sense can it have when one considers the Universe  as a whole? Can we say that one branch of the wave function of the Universe is more probable than another?  Perhaps, it would be more consistent to interprete the wave function of the Universe as a
fundamental object and  to treat all its branches on equal footing? Here arises the idea to combine the many-worlds interpretation of quantum mechanics with Anthropic Principle (see e.g. \cite{Barrow}). Firstly this idea was expressed  in (\cite{Barv1}) and then further developed in (\cite{Kam-Ter}). At first glance, one can think that the many-worlds interpretation is ``anti-anthropic'' in its spirit because it deprives a human being of its privileged position in the Universe. At the same time, it can be combined quite harmoniously with the anthropic principle and cosmology. Indeed, from the point of view of this interpretation the Universe is described by a unique wave function, which is adequate to a quantum reality where all possible versions of the evolution are realized (It is interesting that the fundamental paper by Hugh Everett representing the extended version of his PhD Thesis  was called ``The Theory of the Universal Wave Function'' (\cite{Everett1})!).  In some of the branches of the wave function of the universe not only the classical properties, but also the Life and Mind arise, while in other there is nothing similar. Thus, the world
described by our branch of the wave function exists not because it had to arise because of some necessity, but because it was possible and all the possibilities are realized. It is not necessary to require that our world and other similar worlds are the most
probable from the point of view of the measure on the Hilbert space, where the wave function of the Universe is defined.
It could be just in the opposite way. To be exact, our well structured  part of the unique quantum reality has a low probability while less organized branches are more probable. Such an approach matches well with the Anthropic Principle and common sense. Just like Life is localized in a rather small part of the usual space, it could be localized in a tiny part of the Hilbert space.

However, all of the above does not mean that everything is possible. Indeed, the requirements of consistency of theories can impose rather stringent restrictions on the concrete physical laws which govern dynamics in all possible branches of the wave function of the universe. In the next section we discuss a particular example of this.

\section{Density matrix of the universe and the cosmological bootstrap}

This ia well known that pure states form only a part of possible quantum states of the physical system, and in all of the above examples impurity was a result of decoherence associated with the absence of access of the observer to a certain part of the system. However, what if absence of quantum coherence is fundamentally encoded in the quantum state of the Universe? Moreover, can the dilemma of pure vs mixed state of the Universe be solved at the dynamical level within some minimal assumptions on its initial quantum state? Such an approach was first put forward as Euclidean quantum gravity path integral prescription for the density matrix of the Universe in (\cite{Barv6, Barv7}) and then reinterpreted as microcanonical equipartition in real physical spacetime with the Lorentzian signature (\cite{Barv14, Barv14a}). This model was further developed in (\cite{Barv15, Barv16, Barv17, Barv18,Barv19,Barv20}).

The very idea that instead of a pure quantum state of the universe one can consider a density matrix in the form of the Euclidean quantum gravity path integral was pioneered by D.Page in (\cite{Page}). A corresponding mixed state of the universe arises naturally if there exists an instanton with two turning points (surfaces of vanishing external curvature). As shown in (\cite{Barv6, Barv7}) such an instanton naturally arises and can be dynamically supported if one considers a closed Friedmann universe with the following two essential ingredients: effective cosmological constant and radiation corresponding to the set of conformally invariant quantum fields.
The Euclidean Friedmann equation in this case is
\begin{equation}
\frac{\dot{a}^2}{a^2} =
    \frac{1}{a^2} - H^2 -\frac{C}{a^4},
    \label{dens-Fried}
\end{equation}
where $H^2=\varLambda/3$ is an effective cosmological constant and the constant $C$ characterizes
the amount of radiation in the universe. The turning points for solutions of this equation are
\begin{equation}
a_\pm= \frac{1}{\sqrt{2}H}
    \sqrt{1\pm(1-4CH^2)^{1/2}},\ \ 4CH^2 \leq 1.
    \label{dens-turn}
    \end{equation}
Fig. 1 gives the picture of the instanton which underlies the density matrix of the universe, minimal and maximal values of the oscillating scale factor corresponding to these turning points.
\begin{figure}[t]
\centerline{\epsfxsize 5cm
\epsfbox{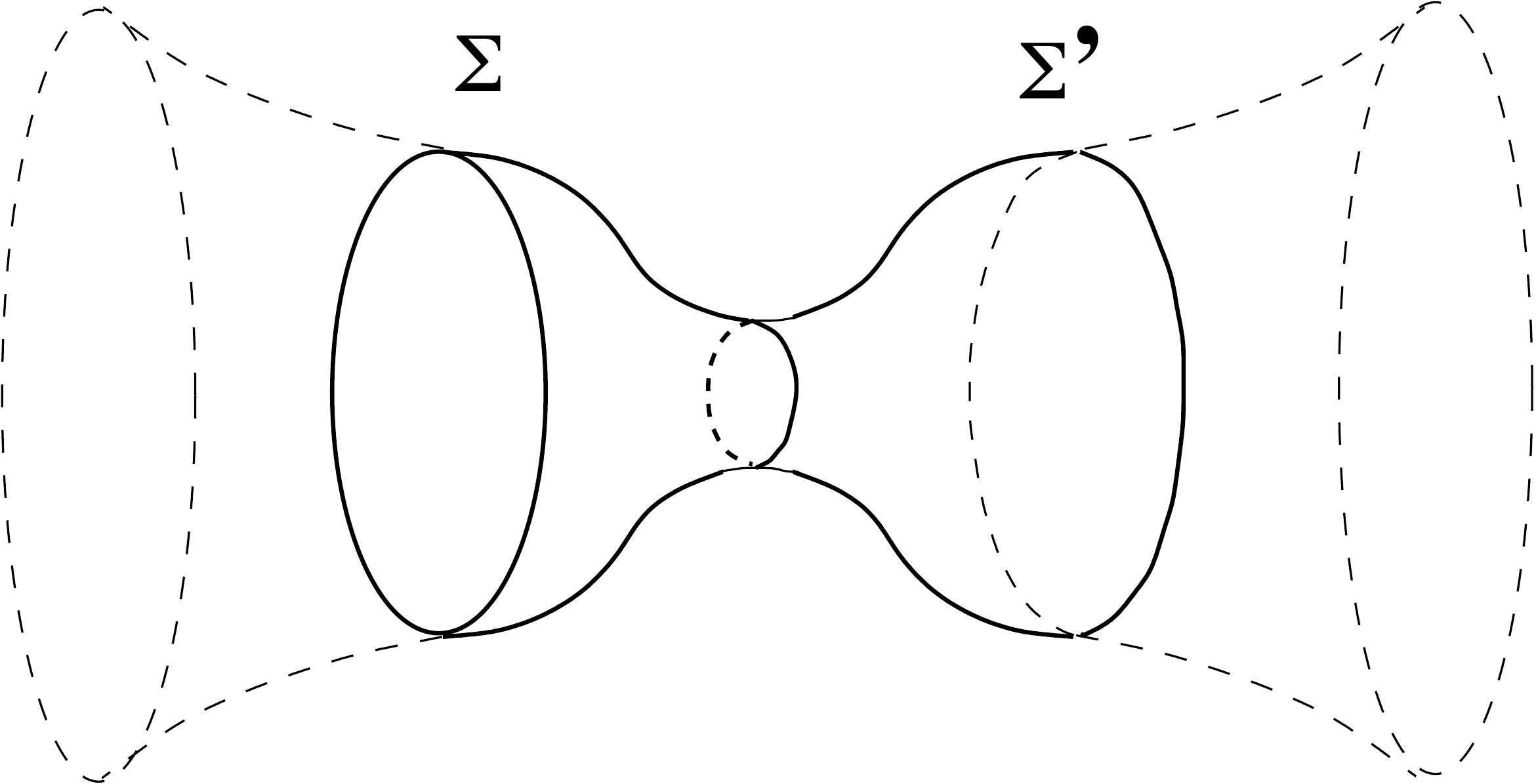}}
\caption{Picture of instanton representing the density matrix. Dashed lines
depict the Lorentzian Universe nucleating from the instanton at the
minimal surfaces $\Sigma$ and $\Sigma'$.}
\end{figure}
For the pure quantum state (\cite{Hawk})  the instanton bridge between
$\Sigma$ and $\Sigma'$ breaks down (see Fig. 2). However, the
radiation stress tensor prevents these half instantons from decoupling -- the minimal value $a_-$ stays nonzero.
\begin{figure}
\centerline{\epsfxsize 5cm
\epsfbox{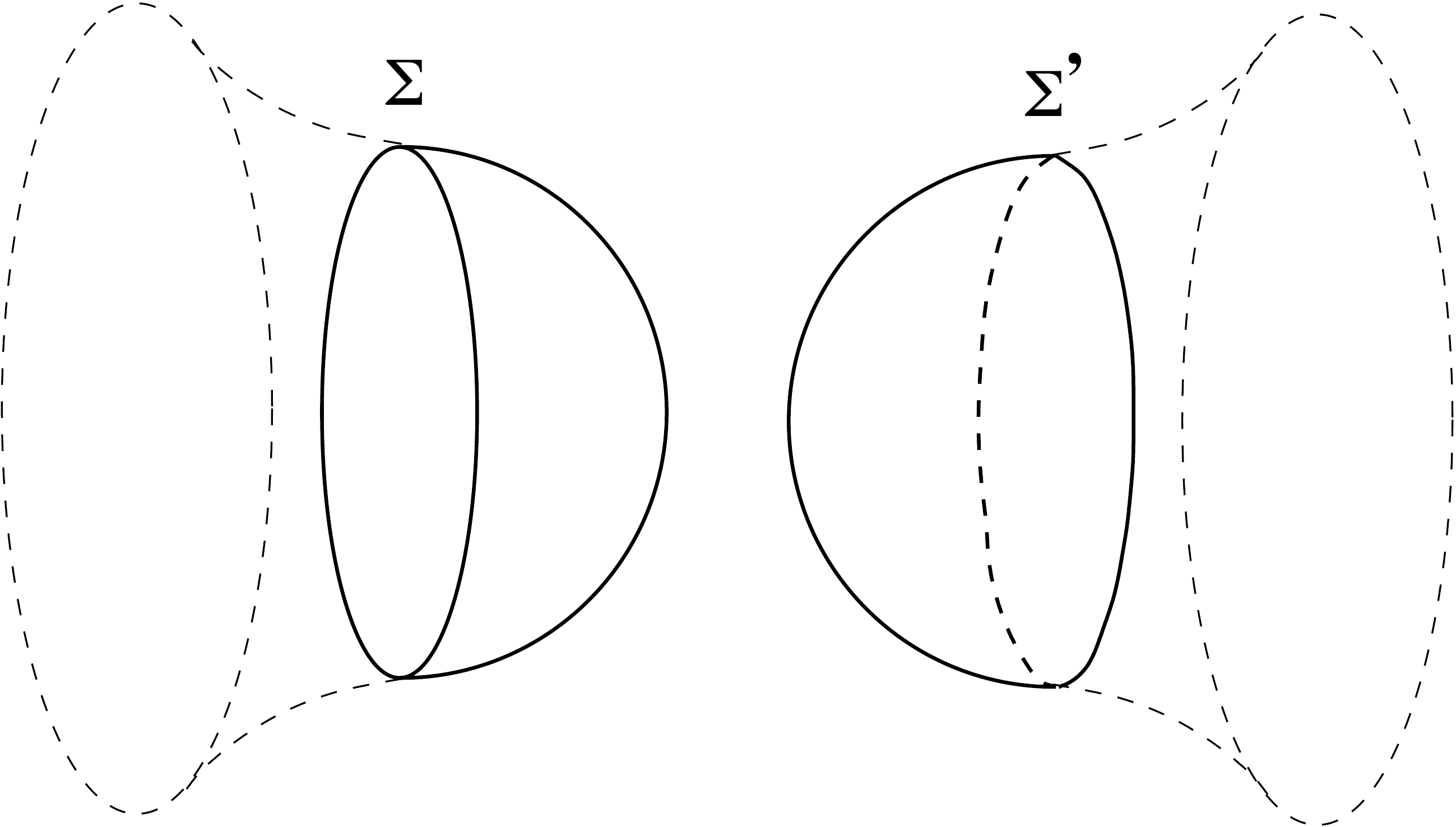}}
\caption{Density matrix of the pure Hartle-Hawking state represented by the
union of two no-boundary vacuum instantons.}
\end{figure}

The relevant density matrix is the path integral over metric and matter field histories interpolating between their boundary values at $\Sigma$ and $\Sigma'$,
\begin{equation}
\rho[\,\varphi,\varphi'\,]=\mbox{$e$}^{
    \varGamma}\!\!\!\!\!\!\!\!\!\!\!\!\!\!
    \int\limits_{\,\,\,\,\,\,\,\,g,\,
    \phi\,\big|_{\;\Sigma,\Sigma'}\,=\,(\,\varphi,\varphi')}
    \!\!\!\!\!\!\!\!\!D[\,g,\phi\,]\,
    \exp\big(-S_{\rm E}[\,g,\phi\,]\big).  \label{DMU}
\end{equation}
Here $S_{\rm E}[\,g,\phi\,]$ is the Euclidean action of the model. The partition function $e^{-\varGamma}$ for this density matrix follows from
integrating out the field $\varphi$ in the coincidence limit of its two-point kernel at $\varphi'=\varphi$. This corresponds to the identification of $\Sigma'$ and $\Sigma$, the underlying Euclidean spacetime acquiring the ``donut'' topology $S^1\times S^3$. The semiclassical saddle point of the path integral for $e^{-\varGamma}$ is just the instanton of the above type.

The metric of this instanton is conformally equivalent to the metric of the Einstein static universe:
\begin{equation}
d{s}^2 = d\eta^2 + d^2\Omega^{(3)},
\label{Ein-stat1}
\end{equation}
where $\eta$ is the conformal time related to the cosmic time $\tau$ by the relation $d\eta=d\tau/a(\tau)$. This opens the possibility of exact calculations for conformally invariant quantum fields, because their effective action on this minisuperspace background is exhausted by the contribution of the conformal anomaly, relevant Casimir energy and free energy. Therefore, at the quantum level the Friedmann equation gets modified to
\begin{eqnarray}
    \frac{\dot{a}^2}{a^2}
    +B\,\left(\frac12\,\frac{\dot{a}^4}{a^4}
    -\frac{\dot{a}^2}{a^4}\right) =
    \frac{1}{a^2} - H^2 -\frac{C}{ a^4},     \label{Friedmann}
    \end{eqnarray}
where the amount of radiation constant $C$ is given by the bootstrap
equation\footnote{We use units with rescaled Planck mass $m_P$ which is related to the gravitational coupling constant $G$ and the reduced Planck mass $M_P$ as $m_P^2=3\pi/4G=6\pi^2M_P^2$.}
   \begin{equation}
    m_P^2 C = m_P^2\frac{B}2 +\frac{dF(\eta)}{d\eta}
    \equiv \frac{B}2 m_P^2+
    \sum_{\omega}\frac{\omega}{e^{\omega\eta}-1}. \label{bootstrap}
    \end{equation}
Here $F(\eta)$ is the free energy which for the conformally coupled scalar field is given by the series of terms contributed by field-theoretical oscillators with frequencies $\omega/a(\tau)$ on a 3-spere of the radius $a(\tau)$
\begin{eqnarray}
&&F(\eta)=\sum_{\omega}
    \ln\big(1-e^{-\omega\eta}\big)
    =\sum_{n=1}^\infty n^2\,
    \ln\big(1-e^{-n\eta}\big).
    \end{eqnarray}
Here $\eta$ is the period of the cosmological instanton in units of the conformal time -- effective inverse temperature of the gas of conformal particles. The constant $B=\beta/8\pi^2M_P^2$ here describes the contribution associated with the conformal anomaly and Casimir energy of the model, where $\beta$ is a dimensionless coefficient of the Gauss-Bonnet term of the stress tensor trace anomaly. Similar expressions hold for other conformally invariant fields of higher spins.

Let us emphasize that we have obtained a highly non-trivial system of equations. While the geometry of the instanton depends on the amount of radiation through the modified Friedmann equation, the amount of radiation, in turn, depends on the parameters
of the instanton. We called this phenomenon ``cosmological bootstrap''.

The Friedmann equation can be rewritten as
\begin{eqnarray}
    &&\dot{a}^2 = \sqrt{\frac{(a^2-B)^2}{B^2}
    +\frac{2H^2}{B}\,(a_+^2-a^2)(a^2-a_-^2)}-\frac{(a^2-B)}{B}          \label{time-der1}
    \end{eqnarray}
and has the same two turning points $a_\pm$ as in the classical case
provided
    \begin{equation}
    a_-^2 \geq B.                \label{require}
    \end{equation}
This requirement is equivalent to
\begin{equation}
    C \geq B-B^2 H^2,\,\,\,\,B H^2\leq\frac12. \label{restriction1}
    \end{equation}
Together with $CH^2 \leq 1/4$ the admissible domain for instantons on a two-dimensional plane of $C$ and $H^2$ reduces to the curvilinear wedge below the hyperbola and above the straight line to the left of the critical point (see Fig. 3):
\[C = \frac{B}{2},\ \ H^2 = \frac{1}{2B}.\]
More detailed analysis shows that cosmological instantons form one-parameter families classified by the number of oscillations of the scale factor during the instanton time period $k=1,2,...$. Because of these oscillations they can be called garlands.
\begin{figure}
\includegraphics[height=.4\textheight]{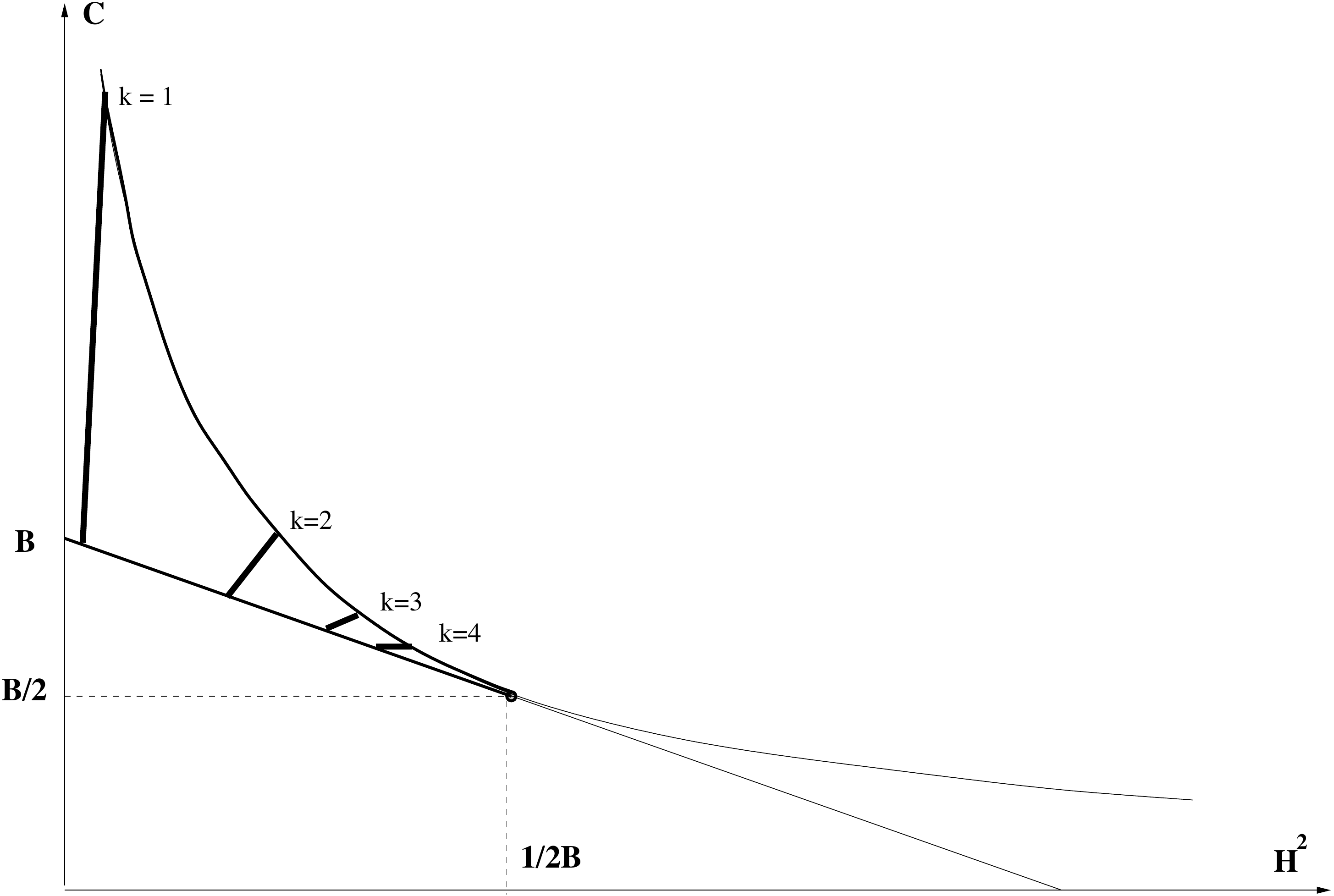}
\caption{The instanton domain in the $(H^2,C)$-plane is located
between bold segments of the upper hyperbolic boundary and lower
straight line boundary. The first one-parameter family of instantons
is labeled by $k=1$. Families of garlands are qualitatively shown
for $k=2,3,4$. $(1/2B,B/2)$ is the critical point of accumulation of
the infinite sequence of garland families.}
\end{figure}

The suggested approach allows one to resolve the problem of the
so-called infrared catastrophe for the no-boundary state of the
Universe based on the Hartle-Hawking instanton. This problem is
related to the fact that the Euclidean action on this instanton is
negative and inverse proportional to the value of the effective
cosmological constant. This means that the probability of the
universe creation with an infinitely big size is infinitely high.
The effect of conformal anomaly allows one to avoid
this counter-intuitive conclusion.

Indeed, outside of the admissible domain for the instantons with two
turning points, obtained above, one can also construct instantons
with one turning point which smoothly close at $a_- = 0$ with $\dot
a(\tau_-)=1$. Such instantons correspond to the Hartle-Hawking pure
quantum state. However, in this case the on-shell effective action,
which reads for the set of solutions obtained above as
    \begin{eqnarray}
    &&\varGamma_0= F(\eta)-\eta\frac{dF(\eta)}{d\eta}
    +4m_P^2\int_{a_-}^{a_+}
    \frac{da \dot{a}}{a}\left(B-a^2
    -\frac{B\dot{a}^2}{3}\right),              \label{action-instanton}
    \end{eqnarray}
diverges to plus infinity. Indeed, for $a_-=0$ and $\dot a_{-}=1$
\begin{eqnarray}
&&\eta = \int_0^{a_+}\frac{da}{\dot a
a}=\infty,\,\,\,F(\infty)=F'(\infty)=0,
\end{eqnarray}
and hence the effective Euclidean action diverges at the lower limit
to $+\infty$. Thus,
\[\varGamma_0 = +\infty,\ \ \exp(-\varGamma_0) = 0,\]
and this fact completely rules out all pure-state instantons,
and only mixed quantum states of the universe with finite values of the effective Euclidean action $\varGamma_0$ turn out to be admissible. This is a dynamical mechanism of selection of mixed states in the cosmological ensemble described by the density matrix.

It turned out that under an appropriate definition of the microcanonical ensemble in the theory of spatially closed universes the picture of the above type can be derived from the canonically quantized gravity theory in physical spacetime with the Lorentzian signature (\cite{Barv14,Barv14a}). The microcanonical density matrix of the Universe, defined as a projector on the subspace of solutions of the Wheeler-DeWitt equation and momentum constraints, can be represented by the same path integral (\ref{DMU}) with the prescription that integration over the Euclidean metric lapse function (in the Arnowitt-Deser-Misner (3+1)-decomposition) should run along imaginary axis. This construction describes a kind of an ultimate equipartition in the physical phase space of the quantum constrained gravity theory. However, in terms of the observable spacetime geometry this equipartition is peaked about a set of cosmological instantons which, according to Fig.3, are limited to a bounded range of values of the cosmological constant $\varLambda=H^2/3$. These instantons obtained above as
fundamental in Euclidean quantum gravity framework, in fact, turn
out to be the saddle points of the gravitational path integral in physical Lorentzian signature spacetime, located at the imaginary axis in the complex plane of the ADM lapse function.

The further development of this concept has shown that this model of initial conditions for a quantum state of the Universe suggests many interesting physical predictions. They include the energy scale of inflation $\varLambda\sim M^2_P/\beta$, which is inverse proportional to the conformal anomaly coefficient $\beta$ (\cite{Barv6,Barv7}), potentially observable thermal imprint on primordial CMB spectrum (\cite{Barv16,Barv19}), new type of the hill-top inflation (\cite{Barv17,Barv18}) in the version of Higgs inflationary model with a large non-minimal inflaton-curvature coupling, etc. In particular, the anticipated hierarchy between the Planck scale and inflation scale suggests a very big value of $\beta$ which could be accessible only in rather popular now theories with numerous higher-spin conformal particles (\cite{Tseytlin}). Moreover, this model undergoes a test on applicability of semiclassical expansion because it stays below a well-known effective theory cutoff (\cite{cutoff,Dvali}), $M^2_P/\beta\ll M_P^2/N$, for large numbers $N$ of gravitating quantum species (\cite{Barv20}).

To summarize this section, we conclude that relaxing a usually tacit assumption of purity for the quantum state of the Universe one can come to nontrivial predictions for basic cosmological parameters, which might lead to resolving such fundamental issues as the cosmological constant problem, the problem of cosmological landscape in large scale structure of the Universe and the others.

\section{Concluding remarks}

We have briefly reviewed  here a series of results hopefully elucidating foundations of quantum theory and various related aspects of quantum cosmology. Some of these results are directly associated with two seminal papers of Heinz-Dieter Zeh. Other results are devoted basically to the problem of the definition of the quantum state of the Universe in quantum cosmology -- the issue which also belongs to H.-D. Zeh scientific legacy. Let us reiterate: one of his works (\cite{measurement}) has laid the foundation of such an approach to quantum theory as decoherence, while the other (\cite{observation}) treated at the technical
level the problem of the preferred basis in the many-worlds interpretation of quantum mechanics at the time when this interpretation was hardly known yet (it has been published the same year when the monograph edited by DeWitt and Graham has just appeared). No doubt that these works of Zeh will attract the attention of researchers also in the future and it is hard to predict the diversity of contexts in which they can produce groundbreaking effect on expansion of our knowledge about Nature.

\begin{acknowledgement}

The work of A.B. was supported by the RFBR grant No 20-02-00297 and by the Foundation for Theoretical Physics Development ``Basis''. The work of A.K. was partially supported by the RFBR grant No 20-02-00411. We are  grateful to
C. Deffayet, C. Kiefer, I.V. Mishakov, D.V. Nesterov, V.N. Ponomariov, O.V. Teryaev, A. Tronconi, T. Vardanyan and  G. Venturi
for useful discussions.

\end{acknowledgement}

 \bibliographystyle{harvard}


\end{document}